\documentclass[12pt,a4paper,preprint]{aastex}

\usepackage{color}
\definecolor{darkblue}{rgb}{0.0, 0.0, 0.4}

\usepackage{hyperref}

\hypersetup{colorlinks=true,citecolor=darkblue,linkcolor=black}

\slugcomment{ApJ, Accepted April 2012}

\shorttitle{Photomerty of Gliese 569B}
\shortauthors{Kenworthy and Scuderi}

\begin{document}

\title{Infrared Variability of the Gliese 569B System}

\author{Matthew A. Kenworthy\altaffilmark{1}}

\affil{Leiden Observatory, Leiden University, P.O. Box 9513, 2300 RA Leiden, The Netherlands}

\and

\author{Louis J. Scuderi\altaffilmark{1}}

\affil{Institute for Astronomy, University of Hawaii, 2680 Woodlawn Drive, Honolulu, HI 96822-1839, USA}

\altaffiltext{1}{Steward Observatory, 933 North Cherry Avenue, Tucson, AZ 85721}

\email{kenworthy@strw.leidenuniv.nl}

\begin{abstract}

Gliese 569B is a multiple brown dwarf system whose exact nature has been
the subject of several investigations over the past few years.
Interpretation has partially relied on infra-red photometry and
spectroscopy of the resolved components of the system. We present seeing
limited $K_s$ photometry over four nights, searching for variability in
this young low mass substellar system. Our photometry is consistent with
other reported photometry, and we report the tentative detection of
several periodic signals consistent with rotational modulation due to spots on
their surfaces. The five significant periods range from 2.90 hours to
12.8 hours with peak to peak variabilities from $28$ mmag to $62$ mmag
in the $K_s$ band.

If both components are rotating with the shortest periods, then their
rotation axes are not parallel with each other, and the rotation axis of
the Bb component is not perpendicular to the Ba-Bb orbital plane. If Bb
has one of the longer rotational periods, then the Bb rotation axis
is consistent with being parallel to the orbital axis of the Ba-Bb
system.

\end{abstract}

\section{Introduction}

Gliese 569B was first discovered and identified as a comoving companion
to nearby low mass star Gliese 569 ($d=9.8$ pc, SpT=dM2) by
\citet{Forrest88}, as part of a survey of nearby stars looking for brown
dwarf candidates \citep{Skrutskie89}. They concluded that the unusual
red colors of this object suggested either a late-type M dwarf, a young
brown dwarf still in the process of contracting, or an unresolved binary
system.  In the twenty years following the initial discovery, the true
nature of this object has been difficult to determine due to the
evolving state of low mass stellar models and the geometry of the
system.

Visible band spectroscopic analysis of the B component \citep{Henry90}
determined that it was similar to several late M-dwarfs and known brown
dwarfs, eventually assigning it a spectral type of M 8.5.  This model
seemed to fit the data relatively well, as it was bluer than a known,
similar age brown dwarf, but redder than a known late M star.  With the
advent of adaptive optics, GL 569B was resolved into two separate
components, GL 569Ba and GL 569Bb \citep{Martin00}. This paper was also
able to constrain the age of the system to between 0.21 Gyr and 1.0 Gyr,
and deduced that the total mass of the system to be $0.09-0.15M_\odot$
because of the observed lithium depletion and IR excess.  They also
estimated the orbital period to be on the order of $\sim 3$ years.

Following the discovery that the GL 569B component is (at least) a
binary system, several papers examined the infrared colors of the system
in J, H, and K. After an initial suggestion by \citet{Martin00}, it was
claimed in \citet{Kenworthy01} that the brighter component of the B
system (GL 569Ba) is itself a double with masses roughly equivalent to
the mass of the Bb component. This was based on the observations that
(i) a blended two component system did not fit the data nearly as well
as the original single M8.5 component, (ii) that the infrared colors
were more consistent with a triple system, and (iii) both H and K band
photometry showed a magnitude difference of 0.7 mag between the two
components on two separate telescopes at two different epochs.

\citet{Lane01} determined the mass function of the resolved components
with a complete orbital solution and a revised age range of 200-500 Myr,
assigning individual masses using relative colors.  However, this paper
found no evidence supporting the claim that the system was a triple,
instead arguing for a double system with one component twice as massive
as the other.  This was based on their observed colors and astrometric
data. \citet{Zapatero04} then confirmed that the Bb component is the
first genuine brown dwarf known without theoretical assumptions and
calculated dynamical masses of the system. However, this group also
found inconsistencies in models, noting that the observed surface
gravity of each component was smaller than predicted by otherwise
consistent evolutionary models. The conclusion reached by this paper,
however, was that the 569B system was a binary.

The relatively short period of the B system ($P\sim 2.4 yr$), combined
with the presence of a bright natural guide star suitable for adaptive
optics assisted observations, has made this a well studied system
\citep{Lane01,Zapatero04,Martin06b,Simon06,Dupuy10,Konopacky10,Femenia11}
Most recently, dynamical orbits for both Gliese 569B about the primary
component A have been reported \citep{Femenia11} and more accurate
orbital determinations of Bb about Ba have been presented in
\citep{Konopacky10,Dupuy10}.

\citet{Simon06} set out to measure the dynamical masses of Gliese 569B
using the orbital motion of the 569B system about the A component.
However, with an earlier age of 100 Myr and  high resolution
spectroscopy of the individual components, they concluded that Gliese
569B was a hierarchical triple brown dwarf system, with all three
components having roughly equal masses of $0.04 M_\odot$, which has been
contested in subsequent literature of this object. Other researchers
have determined masses for the 569B system assuming a two component
model, summarised in Table \ref{masses}.

One possible cause of this seeming discrepancy is that a component in
the brown dwarf system Gliese 569B itself may be variable, and an
indication of this is seen in $K_s$ photometry in \citet{Kenworthy01}
and \citet{Lane01}. The purpose of this study is to determine if the
system shows short term infra-red variability significant enough to
explain the differing relative photometry of $\Delta K_s =0.71\pm 0.11$
in \citet{Kenworthy01} and $\Delta K_s=0.41\pm0.03$ seen by
\citet{Lane01}, and to additionally look for signs of rotational
modulation from the individual components. 

Since Gliese 569B is one of the nearest young multiple brown dwarf
systems, understanding its composition and activity is important in
understanding the formation and evolution of low mass stellar systems.
We describe the observations taken, their reduction and photometry in
Section \ref{obs}. The analysis of the photometry are presented in
Section \ref{phot}, the analysis of this data in Section \ref{dis} with
our conclusions in \ref{conc}.

\section{Observations} \label{obs}

Observations were carried out over 2003 April 20-24 UT using the 1.54m
Kuiper Telescope on Mt. Bigelow.  Data on 2003 April 22 UT was lost due
to weather, with observations made on the remaining four nights.  The
camera \citep{Williams93} uses a NICMOS3 256x256 $40\mu m$ pixel InSb
array, formed from four separate quadrants of 128x128 pixels. Each
quadrant has a separate set of readout electronics, and the quadrant we
use has a gain of 15.3 $e^-$/ADU and a read noise of 73 electrons. The
camera is set at the 0.33''/pixel scale for all nights resulting in a
field of view of 42 by 42 arcseconds. All images are taken using the
$K_s$ filter, which has half-power points at $1.99\mu m$ and $2.32\mu
m$. The photometric calibration we use relies on observations
alternating between the target system and a nearby reference star.  The
field of view of the camera is not large enough to capture both the
target object and the standard star in the same frame, and so each night
of observations consists of two interleaved series of data. Observations
were taken throughout the entire night, but the time resolution of the
data are slightly reduced because of the time required to repoint the
telescope between images.

All night followed both Gliese 569 and the reference star through
transit at an airmass of 1.04, following the objects until sunrise at a
typical airmass of 1.4 to 1.54. All data were fit with a circular
Gaussian function to determine the seeing on each of the four nights -
they were 1.20, 1.38, 1.27 and 0.94 arcseconds respectively, with a
typical variation in measured seeing of 0.18 arcseconds. The mean seeing
over all four nights was 1.20 arcseconds.

Several candidate standard stars were selected from the 2MASS all sky
catalog to be of a similar magnitude to the target system of Gl 569B.
This list was further refined for proximity to the science target,
catalog quality of the photometry and lack of variability as indicated
in the 2MASS catalog. The reference star we use is 2MASS
14545403+1602042 $J=9.706\pm0.020;H=9.249\pm0.019;K=9.144\pm0.020$
\citep{Cutri03}. The spectral type of the standard star is estimated to
be K0-K2 from the visible and near infrared colors, and estimated to be
120pc distant.  Each data frame consists of 30 coadded exposures of 2
seconds duration, resulting in a total on sky integration of 60 seconds
per data frame. The sky background flux is approximately 30 counts at
1.06 airmasses up to 65 counts at 1.50 airmasses. Gliese 569B has a
typical peak value of 250 counts above the sky background level, within
the linear response of the detector. All four nights were clear with no
significant cloud cover. Data reduction is carried out with a
combination of IRAF and PDL scripts.

\begin{figure}[htp]
\centering
\includegraphics[scale=0.8]{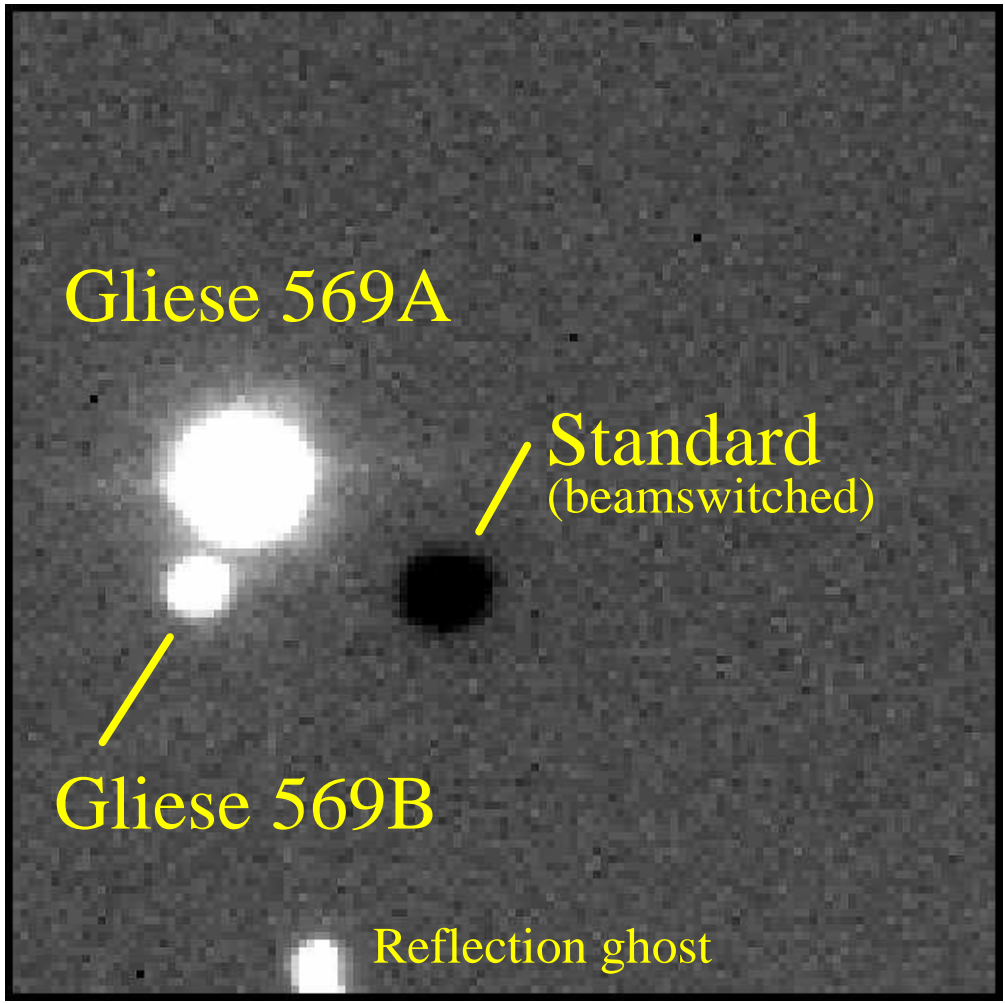}
\caption{Beamswitched image from quadrant [129:256,1:128] of the camera array. The
Gliese 569A star and 569B system are seen as positive flux sources, and
the standard star is shown as the negative flux source. Displayed with a
linear grayscale from -100 to +100 data counts. \label{array}
}

\end{figure}

Dark frames and flat fields were taken with the same exposure time and
readout mode for each night. Twilight flats were taken on the nights of
23 and 24 April. An investigation of the dark frames show a drift in
count levels over a course of integrations, with the drift approximated
by a gradient from 3 to 8 counts across each separate quadrant of the
array between two successive integrations. As a consequence of this
effect, the dark frame and sky background is removed by beam switching.
Each science frame has the closest standard star frame in time
subtracted from it, resulting in a cosmetically clean image with a small
DC component from the change in sky background between frames (see
Figure \ref{array}). The residual sky flux is removed by computing the
median value of a box containing only sky background.

Flat field images were constructed using images of the sky with the
telescope tracking switched off. Dark frames, taken just before the flat
field images, are subtracted off the flat field frames. These dark
subtracted sky images are then combined together with
\textit{IMCOMBINE}, scaling by the mean sky flux and with sigma clipping
to reject any faint star trails that remain in the frame.  The flat
field frames are then normalised by the mean flux within the quadrant
containing the stellar images. These flat fields are divided into the
science observations.  To examine the repeatability of the flat fields
between nights, we produced flat field frames for the nights of 23 and
24 April and divided one into the other. The resultant image shows a
flat image dominated by Gaussian noise with no significant spatial
structure. The flat fields themselves show low spatial structure
gradients in the science region of the images, with no significant flaky
or dead pixels in the region of interest.

\section{Photometry and Analysis}\label{phot}

The slew of the telescope takes approximately ten seconds from target to
reference, and oscillations in the telescope structure take several
seconds to die down to subarcsecond amplitudes. To account for this, a
parameter in the camera control software introduces a fixed delay before
starting the next exposure. In order to maximise our observing
efficiency, we reduced this delay parameter to 6 seconds.  Subsequent
examination of the images showed that in several cases, the oscillations
had not completely died down, leading to an elongation of the stellar
image in the direction of the telescope slew.  The elongation of the
science and reference star images is not consistent, and so we could not
use PSF fitting for data analysis.  Furthermore, the core of the Gliese
569A component was deliberately saturated on the detector, to enable
photometry of the fainter B system. As a consequence of these two
effects, we use aperture photometry for the data reduction.

\begin{figure}[htp]
\centering
\includegraphics[scale=0.7]{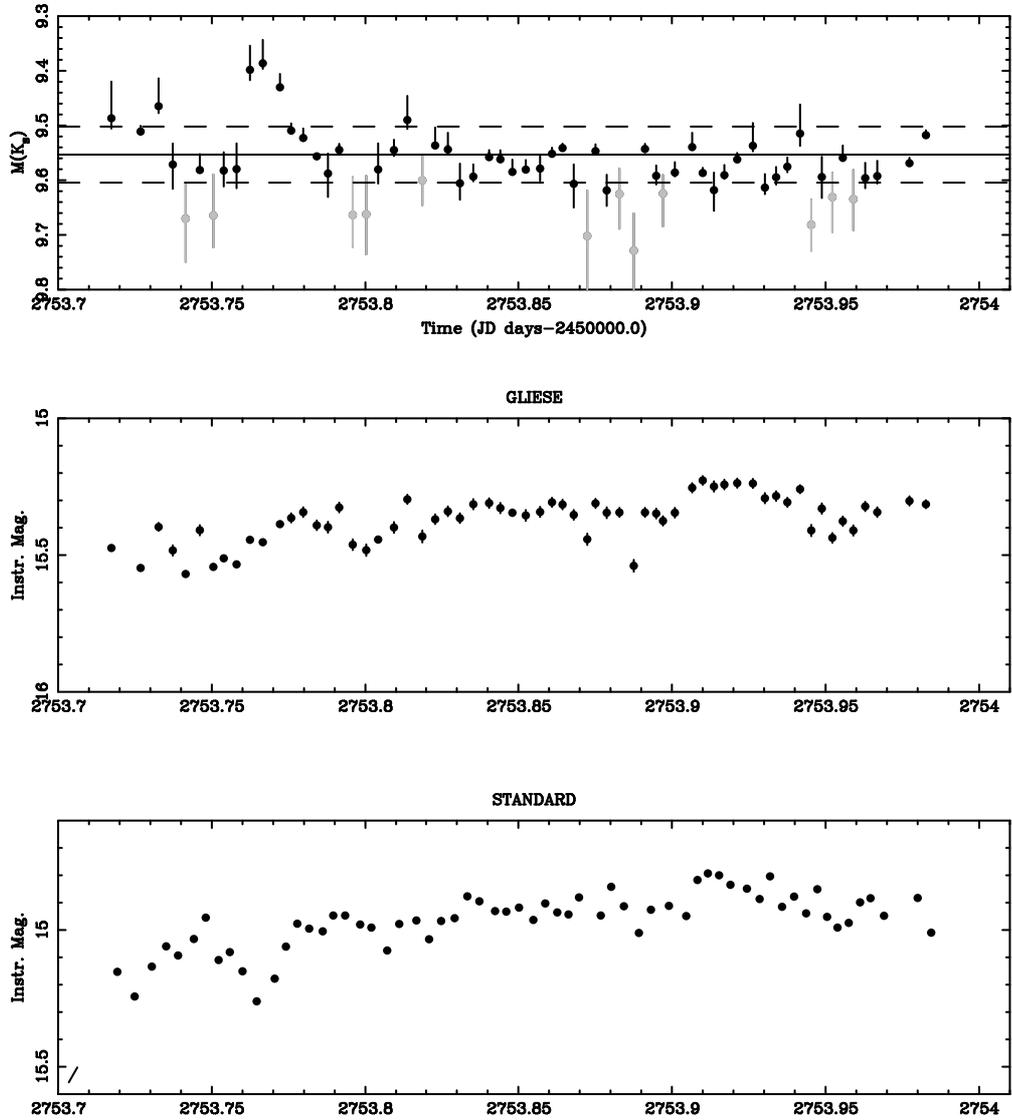}
\caption{The reduced light curve for Gliese 569B on 23 April 2003 (top
panel), with
the uncalibrated photometry for Gliese 569B and the standard star shown
below. Both lower plots show variations of 0.1 to 0.2 magnitudes,
with a slight low order curve attributable to the airmass. The top curve
shows the absolute $K_s$ magnitude of the system. The black dots are
photometry from $r=4$ pixel apertures. The vertical bars on
each point represent aperture photometry for $r=3$ and $r=5$ apertures.
Points with significant variation in aperture photometry are rejected
and are marked in light gray points. The horizonal line represents the mean value
of the black points, and the dashed lines marking the r.m.s. \label{single}}

\end{figure}

Each of the four nights had good natural seeing, allowing for aperture
photometry.  We use a custom PDL routine to identify the Gl569B
component and reference star in all the images. The IRAF \textit{phot}
routine is run on all the data with three extraction apertures, 3.0, 4.0
and 5.0 pixels in radius. Because the telescope had to be repointed each
time an image was taken, the standard star images and the target star
images are taken in an alternating sequence. We linearly interpolate the
instrumental magnitude of adjacent standard star exposures to the time
of the target star observations.  This interpolated standard star
magnitude is then subtracted off the Gl569B magnitude, and this process
is repeated for all three aperture sizes.  The estimated photometric
error is typically $0.020$ magnitudes for the Gl569B frames. This error
is estimated from the noise contributions of the measured sky background
(which includes contributions from the beam switching process, the read
noise and dark current of the detector) and the flux of Gl 569B. This
process is shown in Figure \ref{single}.

Each photometric measurement of GL 569B is calculated  as $M(K_s) = dm
+ m(K_s) - DM$, where $dm$ is the measured magnitude
difference, $m(K_s)$ the K-band 2MASS magnitude of the
standard star, and $DM$ is the correction from apparent to absolute
magnitude for Gliese 569B \citep[$\pi=0.10359\pm 0.00172$;][]{vanLeeuwen07}. The Kuiper
$K_s$ filter is similar in bandwidth to the MKO $K_s$ filter. The
conversions from the 2MASS filter to the MKO K band filter is on the order of
$-0.01$ magnitudes \citep{Dupuy10}. We do not
apply any corrections for the airmass, and we do not apply any color
transformations from the Kuiper telescope filter to the 2MASS $K$ filter
system, as any systematic corrections are on the order of $0.015$ for our
observations.

\begin{figure}[htp]
\centering
\includegraphics[scale=0.7]{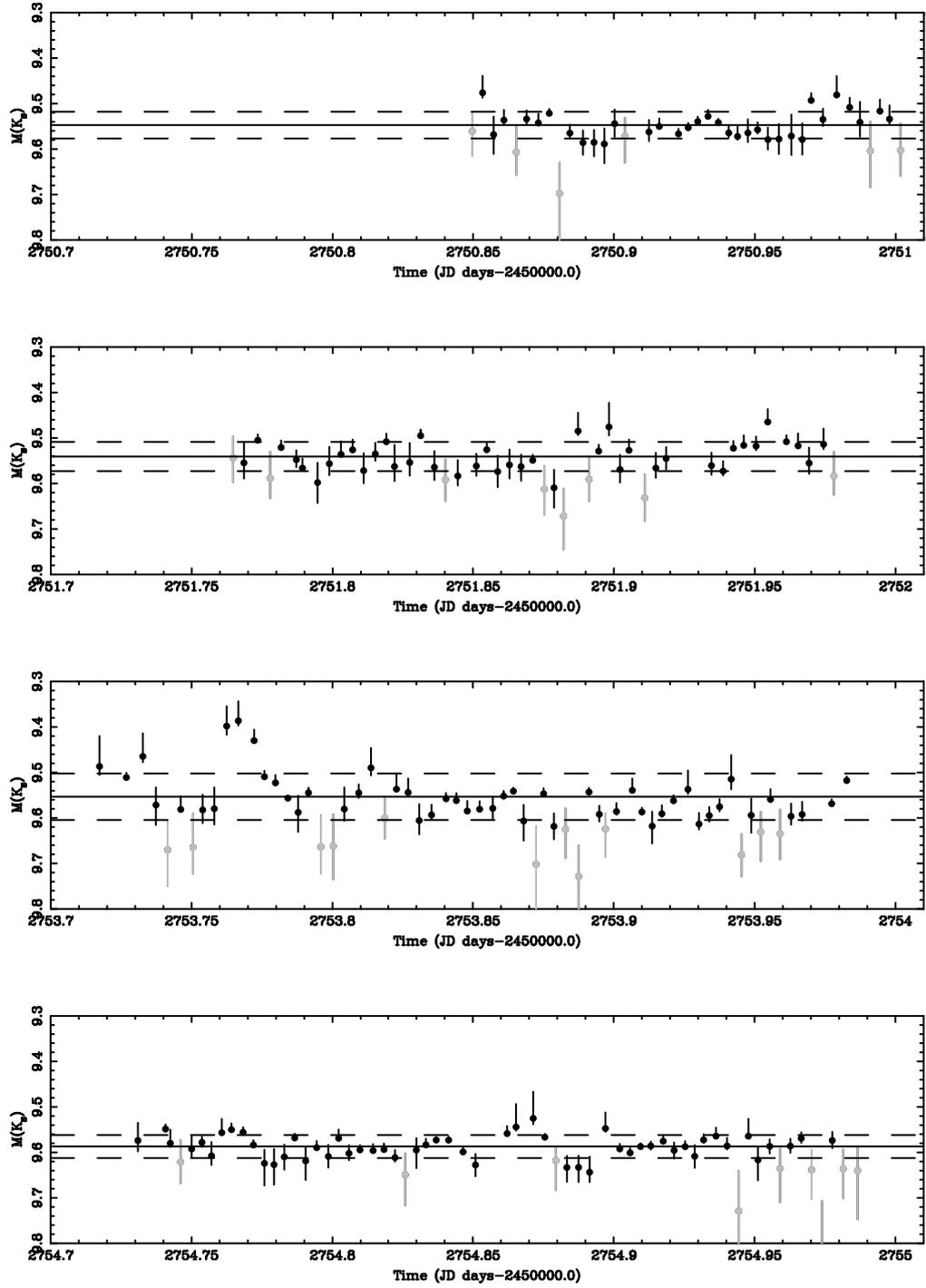}
\caption{\label{multiple}Photometry of Gliese 569B over the dates of
2003 April 20-24 UT. Symbols as in Figure \ref{single}}

\end{figure}

Since the PSF of the standard star and the science target vary due to
the telescope slew, we compared the final differential photometry for
all three aperture sizes. The four nights of data are shown in Figure
\ref{multiple}, where for each data point, the central filled circle
represents the differential photometry for the $r=4$ pixel ($r=1.1$
FWHM) aperture, and the upper and lower ends of the vertical bar
represent the $r=3$ pixel ($r=0.8$ FWHM) and $r=5$ pixel ($r=1.6$ FWHM)
aperture extractions respectively. Ideally, the differential photometry
of all three aperture extractions should be consistent with each other,
within the photometric errors of the individual extractions, which are
on the scale of the diameter of the circles used in Figure
\ref{multiple}. For a majority of the points, the three different
aperture extractions agree with each other to within photometric errors,
but there are on order of 31 points over the four nights which report a
significant spread in extracted photometry. For most of these cases they
are related to manual focus changes in the camera optics, telescope
vibrations, wind shake, and changes in the native seeing over several
minutes. They can also be attributed to the scattered flux from Gliese
569A which is 5 arcseconds away. These points are rejected from further
analysis.

Using the \textit{psfextract} routine to fit a two-dimenstional Gaussian
to the stellar PSFs, we looked for correlations of photometric
variability with the measured FWHM and ellipticity. In several cases we
identified anomalous photometry with poor seeing in either the standard
or target stars. We rejected these points from further consideration and
mark these as grayed circles and lines. Our criterion for rejecting a
data point is one where there is more than 0.1 magnitude spread in
differential photometry between the largest and smallest apertures used in the
extraction.

After the rejection of frames with poor photometry, we are left with an
irregularly sampled set of data covering three timescales - the 24 hour
period of the observations, a seven hour period of the visibility of the
target from the observatory, and a several minute period due to the
alternating observations between reference star and GJ 569B. We look for
sinusoidal periodicities in this irregularly sampled data set by
performing a Lomb-Scargle (LS) periodogram analysis on the photometric data,
as outlined in \citet{Press92} Section 13.8. We construct false alarm
probabilities (FAP) for the LS analysis using the method described in
the previous reference.

\section{Discussion}\label{dis}

\subsection{Short-term Variability from Rotational Modulation}

Both Ba and Bb show broadened absorption lines in infrared
spectra in comparison to theoretical models \citep{Zapatero04}. In a
spectral analysis, \citet{Zapatero04} suggest that this broadening is
due to rotation of the components, and they derive projected rotational
velocities of $v_{rot}\sin i = 37\pm 15\mbox{km.s}^{-1}$ and
$v_{rot}\sin i = 30\pm 15 \mbox{km.s}^{-1}$ for Ba and Bb respectively,
where the uncertainty of the measurements is associated with a poor
knowledge of the molecular opacities used in the model spectra
used as templates. To overcome this,
\citet{Simon06} use observations of Gliese 644C as a template for their
modeling, and they determine projected rotational velocities of  $v_{rot}\sin i
= 25\pm 5\mbox{km.s}^{-1}$ and $v_{rot}\sin i = 10\pm 2
\mbox{km.s}^{-1}$ for Ba and Bb (the errors on the
measurements are given by the velocity intervals of the rotationally
broadened templates - Simon, priv. commun.), where the analysis was
carried out at H band in the spectral orders 48 and 49 of
NIRSPEC. A more recent analysis by \citet{Konopacky12} using K band
spectra with NIRSPEC and synthetically generated spectra using the
PHOENIX atmosphere models yields projected rotational velocities of
$v_{rot}\sin i = 19\pm 2\mbox{km.s}^{-1}$ and $v_{rot}\sin i = 6\pm 3
\mbox{km.s}^{-1}$ for Ba and Bb respectively. The \citet{Konopacky12}
measurements agree with the \citet{Simon06} measurements at the
$1\sigma$ level, but show a systematically smaller projected rotation
velocity. In the subsequent analysis, we use the combined weighted
measurements from both \citet{Konopacky12} and \citet{Simon06}, where 
$v_{rot}\sin i = 19.8\pm 1.9\mbox{km.s}^{-1}$ and $v_{rot}\sin i =
8.8\pm 1.7 \mbox{km.s}^{-1}$ for Ba and Bb.

\begin{figure}[htp]
\centering
\includegraphics[angle=270,scale=0.8]{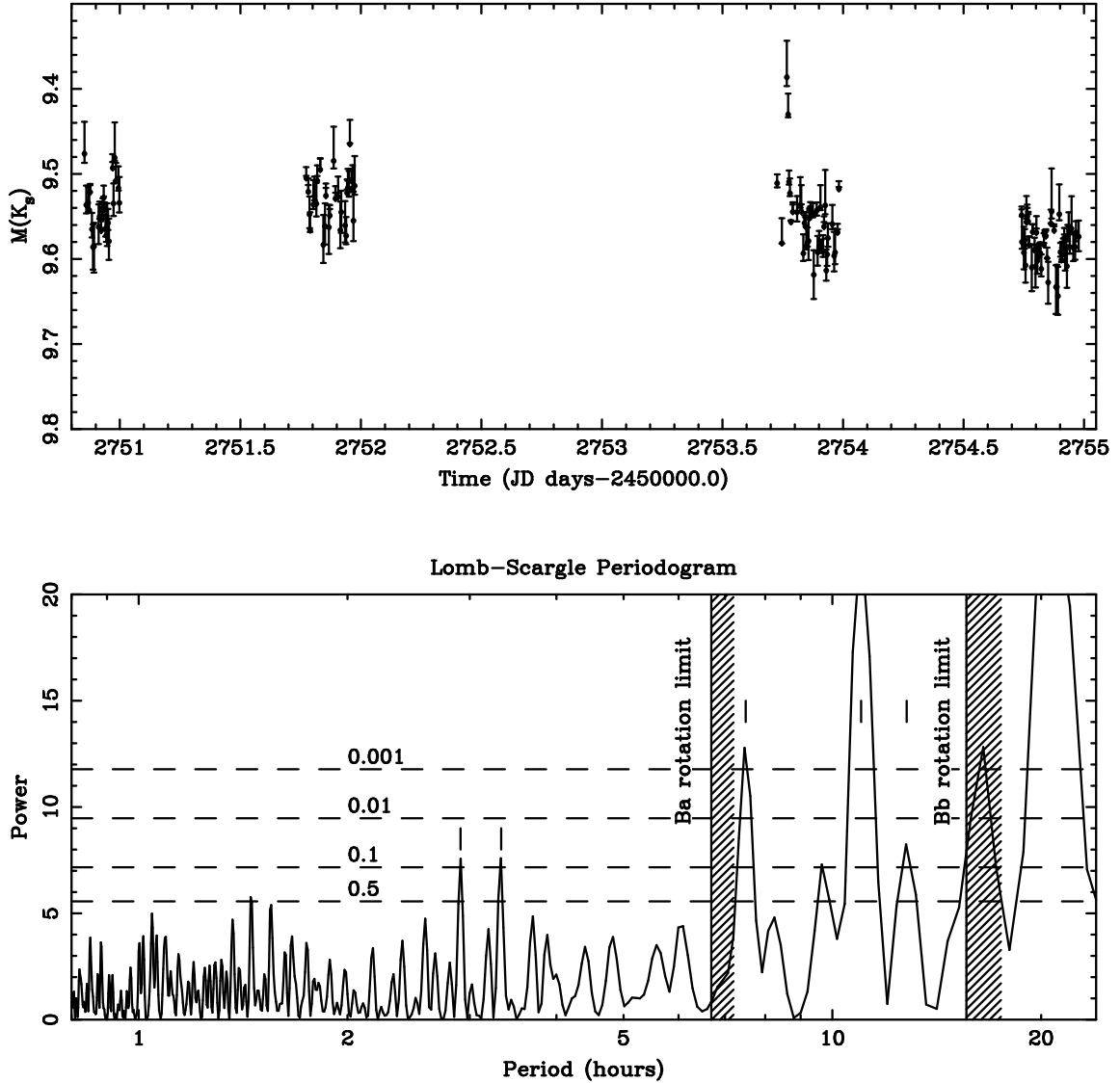}
\caption{\label{ls}Time series phototometry and Lomb Scargle periodogram of Gliese
569B. The upper panel shows the time series photometry, and the lower
panel shows the Lomb Scargle periodgram for the data. The horizontal
lines indicate different levels of False Alarm Probability for the
detected power. The shaded regions indicate the longest allowed
rotational periods given the projected velocities for Ba and Bb.} 

\end{figure}

For a range of plausible models of stellar radii, limited by
uncertainties in the age of the stars, \citet{Zapatero04} pointed out
that the rotational periods of Ba and Bb would be on the order of $3\sin
i$ and $5 \sin i$ hours respectively.
More specifically, given a radius of $0.11R_\sun$, the
rotational periods (in hours) are $6.7 \sin i_{Ba}$ and $15.2 \sin
i_{Bb}$ where $i$ is the angle of inclination between the rotation axis
of the object and our line of sight. Any rotational modulation should be
present at periods shorter than these periods, and these candidate
rotational periods are indicated with vertical lines in Figure \ref{ls}.
The Lomb-Scargle periodogram in Figure \ref{ls} shows two equal power
peaks with false alarm probabilities (FAP) of $<$10\% at $3.320\pm
0.035$ hours and $2.905\pm0.030$ hours (where we have determined the
errors on the periods as the full width at half power in the
periodogram) indicated with tick marks in the figure. These are candidate periods for
both Ba and Bb. Although the individual peaks have relatively large FAPs
associated with them, these are the two most significant peaks over
plausible rotational periods for both Ba and Bb with nearly identical
amounts of power present. We looked for these periods in subsets of the
complete time series, but it requires at least a three day coverage to
provide the temporal resolution in the Lomb-Scargle periodogram for
these two periods. We see these two periods at a lower significance when
we remove the data from either the first day or the last day from our
analysis.

Since Bb has a lower projected broadening velocity, a larger range of
periods can be considered, as indicated in Figure \ref{ls} with
signficant periods at $7.5\pm0.1$ h, $11.0\pm0.5$ h and $12.8 \pm 0.7$ h. The
errors on the period determination are noticably broader as these
periods cover a larger fraction of the total photometric time series
measured.

\begin{figure}[htp]
\centering
\includegraphics[angle=270,scale=0.55]{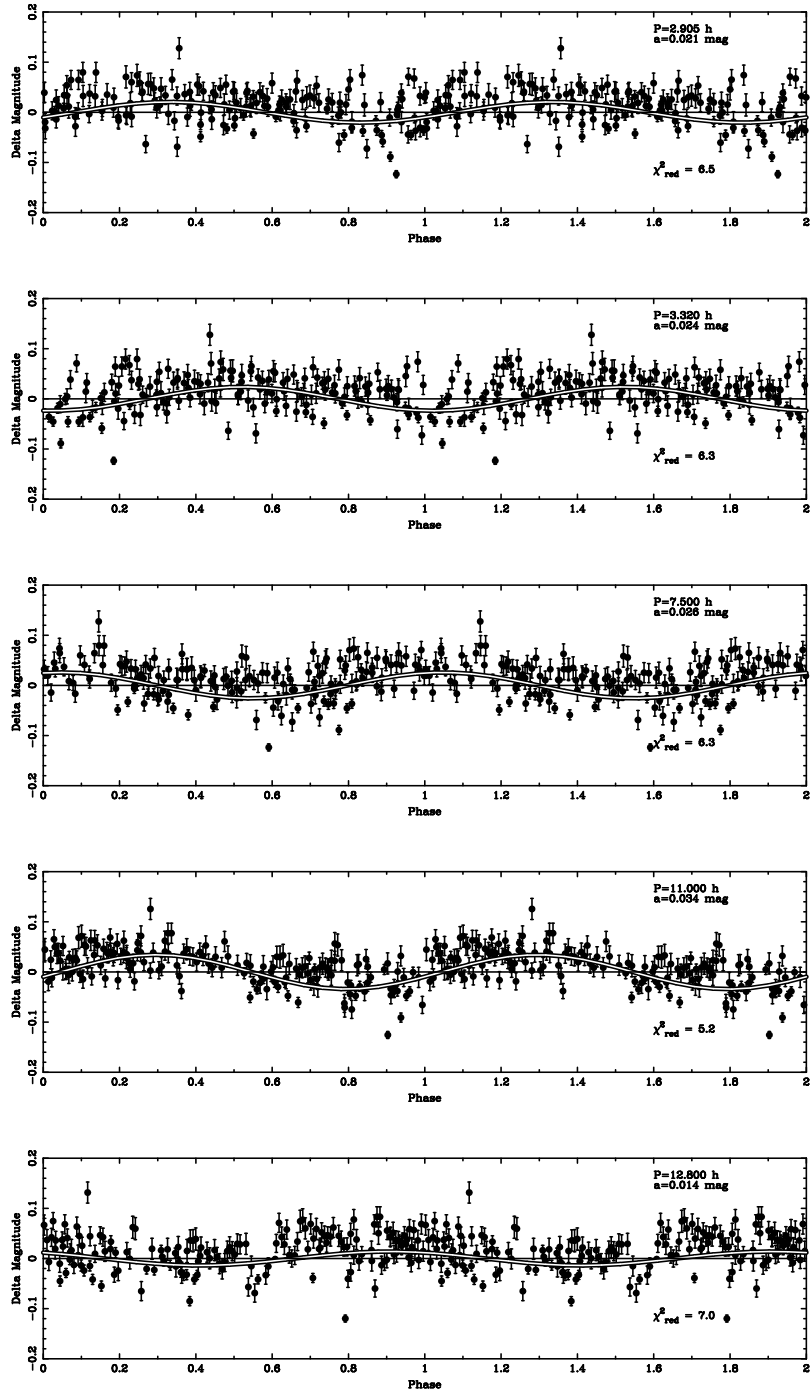}
\caption{\label{folded}Folded light curves of the Gliese 569B photometry
for the periods listed in Table \ref{amplitudefits}. The phase coverage is extended
over two periods for clarity. The best fit sinusoid is drawn over the
folded light curve with the best fit amplitude indicated in the figure
panels.}

\end{figure}

Folded light curves for all five periodicities, along with the best
sinusoidal curve fits are shown in Figure \ref{folded}. All the light
curves show departures from an exact sinusoidal variation. The peak to
peak amplitude of these fluctuations are listed in Table
\ref{amplitudefits}. We determined the
errors on the amplitudes of the fitted sinusoidal functions by
increasing the amplitude of the fitted period and calculating the full
$\chi^2$ value. When this chi-squared value had increased by the mean
value of the reduced chi-squared, we use the delta in fitted amplitude
as an estimate for the error in the fit.  For Ba, the most significant
periods are the 3.320 and 2.905 hours, but because of the lower
rotational broadening seen in Bb, there are additionally significant
periods and amplitudes at 7.5, 11 and 12.8 hours for Bb only.

Variability on these time scales and at these amplitudes is seen in
other late M/early L stars, such as BRI 0021-0214 \citep{Martin01} and
young stars such as S Ori 45 \citep{Zapatero03}. Indeed, variability is
seen with peak to peak values of 30 to 60 mmag \citep{Lane07,Artigau09}
and in the $K_s$ band, \citet{Enoch03} see variability with an amplitude
of 0.2 magnitudes for objects at the L/T transition. The longest period
consistent with the hypothesis of stellar spot modulation shows a
slightly higher $\chi_{red}^2$ goodness of fit, and a significantly
lower peak amplitude compared to the shorter period amplitudes.  There
is also power seen at longer periods which, if real, cannot be due to
rotational modulation in Ba. We attribute these to slowly evolving dust
cloud features \citep{Martin01,Bailer-Jones01}, but simultaneous
observations at other wavelengths are required to test the origin of
these variations.

We emphasise that we cannot uniquely determine what components are
responsible for the periodic signals we are detecting from this seeing
limited data set. We can, however, model what the derived inclinations
would be for different combinations of periods with the Gliese 569 B
components.

\subsection{Derived Inclinations of the Rotational Axes of Gliese 569Ba and Bb}

In order to see what the inclination angle of the rotational axes for Ba
and Bb are, we solved the equation $\sin i = (v_{rot} \sin i).P / (2\pi
R)$ for $i$, where $P$ is the rotational period of the low mass
component and $R$ is the radius of the low mass component. To propagate
the errors in the measurements of $v_{rot}.\sin i$, $P$ and $R$, we run
a Monte Carlo simulation of 100000 runs where values for each of the
input parameters are drawn from normal distributions with sigmas equal
to their respective quoted errors. The resultant distributions of the
inclination angle are then constructed and normalised to the peak value.

\subsubsection{Ba and Bb both have short rotational periods}

\begin{figure}[htp]
\centering
\includegraphics[angle=270,scale=0.8]{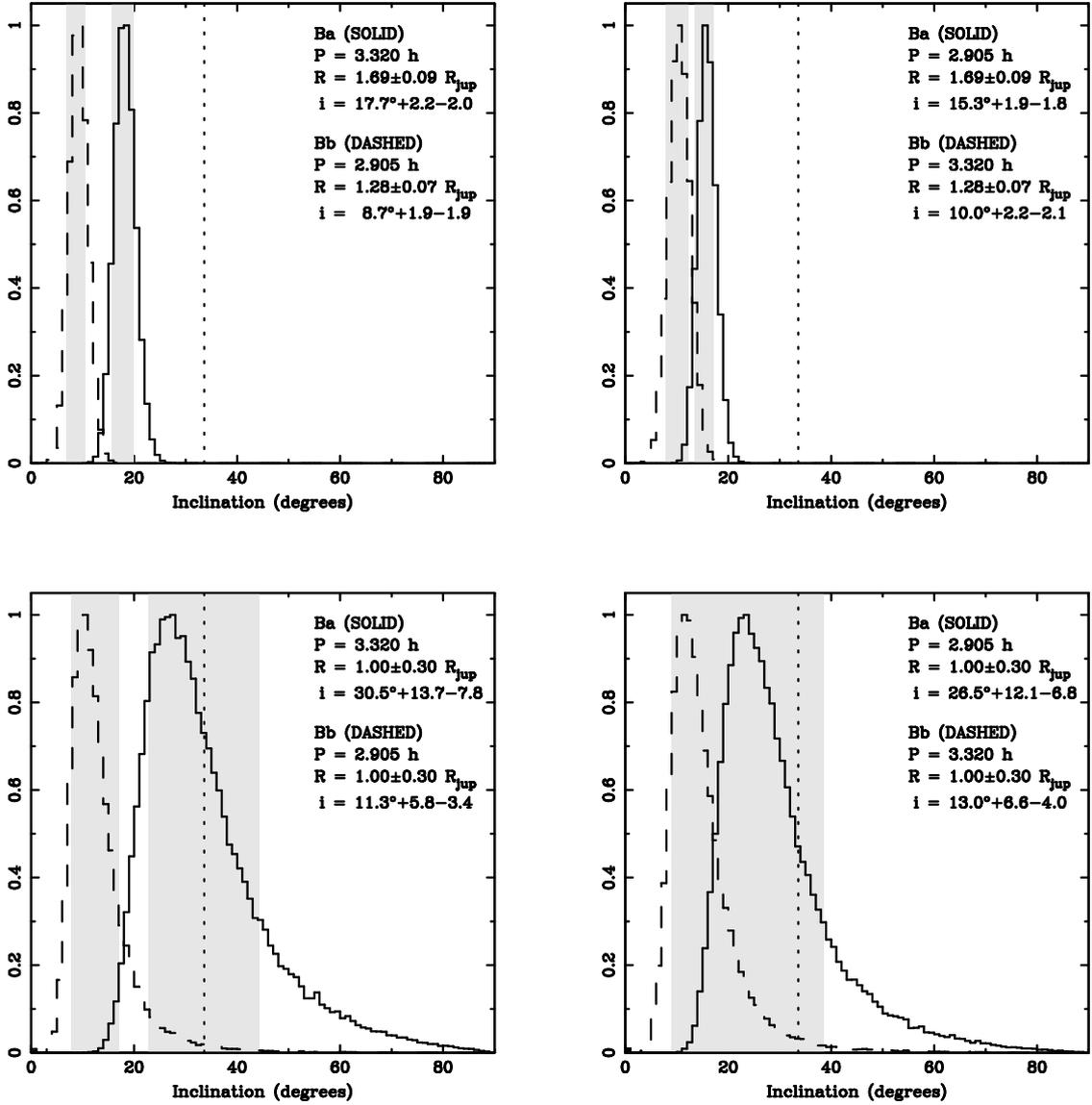}
\caption{\label{tilts}Peak normalised distributions of the rotational
axis inclinations of Ba and Bb from Monte Carlo simulations. Ba
distributions are solid line histograms, Bb are the dashed line
histograms. The confidence limits for each distribution are shown as a
gray box covering 15.9\% to 84.1\% cumulative limits. The vertical
dashed line is the inclination of the perpendicular of the Ba-Bb orbital
plane to our line of sight. The upper two panels are for PHOENIX model
parameters, and the lower two panels cover an extended range of generic
brown dwarf parameters. Each plot represents different input
parameters explained in the text and indicated on the figures.}

\end{figure}

We first consider the case where both Ba and Bb have rotational periods
shorter than 6.5 hours. We are motivated to do this because the FAP of
the two most significant periods are approximately equal in size.  The
resultant distributions of $i$ for each component are shown in Figure
\ref{tilts}.  It is not known which rotational period belongs to Ba or
Bb, so we perform one set of simulations with Ba having the longer of
the two periods, and then we rerun the simulations with the two periods
swapped between the two components. The resultant distributions in
rotational axis inclination are shown as histograms in Figure
\ref{tilts}. We quote the measurement and uncertainties in inclination
angle as confidence limits at 15.9\% and 84.1\% completeness of the
distributions (equivalent to $1 \sigma$ limits in a Normal distribution)
and these are represented as the dark grey boxes on top of each
distribution. The dotted vertical line in all the plots represents the
inclination of the perpendicular to the orbital plane of the Ba-Bb
system \citep[$i = 33.6^o$;][]{Konopacky10}.

The radii of Ba and Bb are estimated in \citet{Konopacky10} by fitting
PHOENIX models to measured photometry. In the upper row of plots in
Figure \ref{tilts}
we use these values and errors on the radii as input to our simulations,
$R_{Ba} = 1.69 \pm 0.09 R_{jup}$ and $R_{Bb} = 1.28 \pm 0.07 R_{jup}$.
The resultant distributions for $i_{Ba}$ and $i_{Bb}$ show that both Ba
and Bb have their rotational axes tilted with respect to one another by
approximately 10 degrees, regardless of which component has the faster
rotation period. Furthermore, both rotational axes are significantly
tilted with respect to the perpendicular of their mutual orbital plane,
i.e. the spin-orbit inclination $i_{rel}$ is non-zero.

To see how robust this result was, we perform a second set of
simulations where we expand the radii of both components to cover a much
wider range of possible radii (including the $0.11R_\odot$ radius used
earlier) - $R_{Ba} = R_{Bb} = 1.0 \pm 0.3 R_{jup}$
and the results form the lower pair of plots in Figure \ref{tilts}. In this
case, both distributions of axial inclination are broader (which is to
be expected) and Ba's rotation axis may be consistent with the orbital
inclination. However, the relative tilt of the rotational axes with
respect to each other remains significant.

\subsubsection{Bb has a longer rotational period}

\begin{figure}[htp]
\centering
\includegraphics[angle=270,scale=0.55]{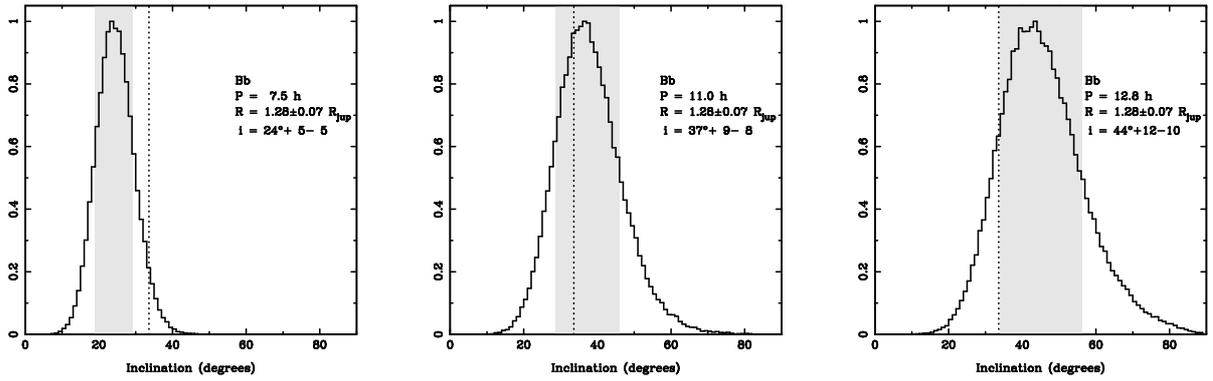}
\caption{\label{tiltslong}Peak normalised distributions of the rotational
axis inclinations of Bb from Monte Carlo simulations. The confidence limits for each distribution are shown as a
gray box covering 15.9\% to 84.1\% cumulative limits. The vertical
dashed line is the inclination of the perpendicular of the Ba-Bb orbital
plane to our line of sight. Each plot represents different input
parameters explained in the text and indicated on the figures.}

\end{figure}

We now consider if Bb has the longer rotational period at 7.5, 11 or
12.8 hours. We run a set of Monte Carlo simulations assuming the same
radius for Bb as derived from the PHOENIX models in the previous
section. The resultant inclination angle distributions are shown in
Figure \ref{tiltslong}.  For 11.0 and 12.8 hours, the inclination of Bb
is consistent with a spin-orbit inclination of zero, although at a lower
level of confidence than with the shorter periods in the previous
section. At the shortest of the three periods, there is evidence that Bb
may have a spin-orbit inclination of approximately ten degrees.

\section{Conclusions}\label{conc}

We draw three main conclusions from our data and analysis.

The data presented here cover four days of photometric monitoring, where
there is no variability in the Gliese 569B system large enough to
explain the difference in relative brightness at $K_s$ between Ba and Bb
seen in \cite{Kenworthy01} and \cite{Lane01}. Our first conclusion is
that there is an unaccounted systematic error in the data reductions of
one or both groups. The photometry we present in this paper is
consistent with other measurements in the literature \citep[e.g. see
][]{Forrest88,Martin00,Kenworthy01,Lane01,Simon06,Dupuy10}.

Analysis of the time series of data presented in this paper show
variability in the Gliese 569B system above that expected for
measurement noise alone. Our analysis shows the presence of five
periodicities from 3.32 to 12.8 hours with a peak to peak
amplitude of 0.04 to 0.06 magnitudes, consistent with variability seen
in other young low mass star systems.

If we attribute the periodicity in the light curve to rotational
modulation of starspots on Ba and Bb, we can draw two further
conclusions. Assuming Ba is a single object and that one of the two
shorter periods is from Ba, then the rotational axis of Ba is not
parallel to the orbital axis of the Ba-Bb system for fitted PHOENIX
models of the radius of Ba.  If this result is confirmed independently,
it will be interesting to determine whether this spin-orbit misalignment
is primordial in nature or due to subsequent interactions from the
third, more massive 569 A component. Currently few stellar examples
exist, although misaligned disks in triple systems have also been
detected \citep{Skemer08}. To determine the contribution of 569A,
further astrometry is required.  Currently the orbital period of A about
B is estimated to be approximately 400 years, with a low ellipticity
constraint \citep{Femenia11}.

If the Ba and Bb components are responsible for the two shortest
periods, then we can robustly say that the Ba and Bb rotational axes are
not aligned with respect to each other. Alternatively, if Bb is
responsible for one of the 7.5 or longer periods, then Bb's rotation axis is
aligned with the orbital axis of the Ba-Bb system. The conclusion that
Ba's rotational axis is inclined to the spin-orbit axis remains valid.

Future work should include a longer and more sensitive
photometric observing campaign to confirm the rotational modulation.
More long term monitoring will begin to investigate the shape of the
modulated light curve and determine if we are seeing the effects of
clouds in the atmospheres of this multiple brown dwarf system.  Longer
monitoring with an adaptive optic telescope will allow splitting of the
components and confirmation of the separate rotational periods. Higher
spectral resolution measurements in the individual 569B components will
begin to constrain the theoretical models of these objects, assuming the
spectral line broadening is due to stellar rotation.

Gliese 569B is a benchmark system whose proximity and bright
components will lead to further investigations and act as an ongoing
test for low mass stellar models.

\bibliographystyle{apj}                       
\bibliography{apj-jour,kenworthy}

\acknowledgments

{\it Facilities:} \facility{Kuiper 61 inch Telescope}.

We thank the anonymous referee for many useful comments and suggestions
which have improved this paper and for their patience with these
revisions. We thank Marcia and George Rieke for their support of the
$256^2$ IRCAM that made these observations, and to Chad Engelbracht for
many useful discussions on infra red arrays and the data processing
required. MAK and LS were supported by grant NNG 06-GE26G from the NASA
Terrestrial Planet Finder Foundation Science Program. This research used
the SIMBAD database, operated at CDS, Strasbourg, France, and data
products of 2MASS, which is a joint project of the University of
Massachusetts and IPAC at the California Institute of Technology, funded
by NASA and NSF, and NASA's Astrophysical Data System.  \clearpage

\begin{deluxetable}{llll}
\tablecaption{Masses determined for the Gliese 569B system as reported
in the
literature. Data 
from Table 7 in \citet{Femenia11}, where corrections for the revised
parallax of the 569B system \citep{vanLeeuwen07} have been applied.\label{masses}}
\tablewidth{0pt}
\tablehead{\colhead{Paper} & \colhead{Gliese 569B mass $(M_\odot)$} &
\colhead{Ba $(M_\odot)$} & \colhead{Bb $(M_\odot)$} }
\startdata
\citet{Zapatero04}  & $0.119\pm0.005$             & $0.071\pm0.011$ & $0.054\pm0.011$ \\
\citet{Simon06}     & $0.119\pm0.007$             & $0.100\pm0.011$ & --- \\
\citet{Konopacky10} & $0.120\pm0.007$             & $0.073\pm0.008$ & $0.053\pm0.006$ \\
\citet{Dupuy10}     & $0.140^{+0.009}_{-0.008}$   & $0.075\pm0.004$ & $0.065\pm0.004$ \\
\citet{Femenia11}   & $0.116\pm0.007$             & $0.081\pm0.010$ & $0.059\pm0.007$
\enddata
\end{deluxetable}

\clearpage

\begin{deluxetable}{lll}
\tablecaption{Periods and fitted sinusoidal amplitudes from the Gliese
569B light curve\label{amplitudefits}}
\tablewidth{0pt}
\tablehead{\colhead{Period (h)} & \colhead{Peak to Peak Amplitude (mmag)}  &
\colhead{ $\chi_{red}^2$ } }
\startdata
$3.320\pm0.035$   &   $48\pm 8$     & 6.5 \\
$2.905\pm0.030$   &   $42\pm 8$     & 6.3 \\
$7.5\pm0.1$       &   $52\pm 10$    & 6.3 \\
$11.0\pm0.5$      &   $68 \pm 8$   & 6.2 \\
$12.8\pm0.7$      &   $28 \pm 10$   &   7.0 \\
\enddata
\end{deluxetable}

\clearpage

\clearpage

\clearpage

\clearpage

\clearpage

\clearpage

\clearpage

\end{document}